\documentclass[12pt]{article}
\usepackage{amsmath,amssymb,theorem,graphicx,color}

\usepackage[all]{xy}

\usepackage{color}
\definecolor{dblue}{rgb}{0,0,.7}
\definecolor{indigo}{RGB}{50,0,105}

\usepackage[dvips, hyperindex=true,
colorlinks=true, linkcolor=dblue,
citecolor=indigo, pagecolor=dblue,
bookmarksopen=false, urlcolor=dblue,
linktocpage]{hyperref}

\def\P{{\mathbb P}} \def\O{\mathcal{O}}
\def\C{{\mathbb C}}

\def\CY{Calabi-Yau} \def\c#1{\mathcal{#1}}
\def\R{{\mathbb R}} 
\newcommand{\be}{\begin{equation}}
\newcommand{\ee}{\end{equation}}
\newcommand{\eq}[1]{Eq.~(\ref{#1})}
\def\H{\operatorname{H}}

\newcommand{\bbc}{{\mathbb C}}

\newcommand{\Aut}{{\operatorname{Aut}}}
\newcommand{\vol}{{\text{vol} }}

\newcommand\bbeta{{\bar{\beta}}}

\newcommand\bs{{\bar{s}}}

\newcommand{\vev}[1]{\langle{#1}\rangle}
\newcommand{\cp}[1]{{\mathbb P}^{#1}}

\newcommand{\cL}{{\mathcal L}}

\newtheorem{theo}{Theorem}[section]

\newtheorem{prop}[theo]{Proposition}

{\theorembodyfont{\rmfamily}
}

\def\pplogo{\vbox{\kern-\headheight\kern
-29pt
\halign{##&##\hfil\cr&{\ppnumber}\cr\rule{0pt
}{2.5ex}&\ppdate\cr}}}
\makeatletter
\def\ps@firstpage{\ps@empty
\def\@oddhead{\hss\pplogo}%
  \let\@evenhead\@oddhead}
\def\maketitle{\par
 \begingroup
 \def\thefootnote{\fnsymbol{footnote}}
 \def\@makefnmark{\hbox{$^{\@thefnmark}$\hss}
}
 \if@twocolumn
 \twocolumn[\@maketitle]
 \else \newpage
 \global\@topnum\z@ \@maketitle
\fi\thispagestyle{firstpage}\@thanks
 \endgroup
 \setcounter{footnote}{0}
 \let\maketitle\relax
 \let\@maketitle\relax
 \gdef\@thanks{}\gdef\@author{}\gdef\@title{}
\let\thanks\relax}
\makeatother

\begin{document}
\setcounter{page}0
\def\ppnumber{\vbox{\baselineskip14pt
\hbox{hep-th/0612075}}}
\def\ppdate{RUNHETC-2006-32} \date{}

\title{\bf \LARGE 
	Numerical Calabi-Yau metrics
\\[10mm]}
\author{\bf 
     {Michael R. Douglas$^\&$,
     Robert L.~Karp, 
     Sergio Lukic}\\
     {\bf and Ren\'e Reinbacher} \\
[10mm]
\normalsize  Department of Physics, Rutgers
University \\
\normalsize Piscataway, NJ 08854-8019  USA
		\\	[10mm]
\normalsize $^\&$ I.H.E.S.,  Le Bois-Marie,
Bures-sur-Yvette, 91440 France}

{\hfuzz=10cm\maketitle}

\vskip 1cm

\begin{abstract}
\normalsize
\noindent
We develop numerical methods for approximating Ricci flat metrics on
Calabi-Yau hypersurfaces in projective spaces.  Our approach is based
on finding balanced metrics, and builds on recent theoretical work by
Donaldson.  We illustrate our methods in detail for a one parameter
family of quintics.  We also suggest several ways to extend our
results.
\end{abstract}

\vfil\break

\tableofcontents

\section{Introduction}    \label{s:intro}

Calabi-Yau manifolds were proven to admit Ricci flat metrics in
\cite{Yau}.  Explicit expressions for such metrics would have many
applications in mathematics and in string compactification; however it
is widely thought that for compact Calabi-Yau manifolds
no closed form expression exists,
except in trivial cases (tori and orbifolds).
Thus it is useful to develop methods for constructing and working with
approximate Ricci-flat metrics; and for the solutions of related
equations, such as hermitian Yang-Mills, and the other equations of 
string compactification.

One such approximating method is finding the balanced metric on an
embedding of the manifold into complex projective space,
given by the sections of a holomorphic line bundle.
This method was
suggested by Yau in the early 90's, and was proven to work in a
fundamental paper by Tian \cite{Tian:metrics}. Since then many people
have worked on the problem of balanced metrics in various contexts
\cite{Bourguignon:Yau,Luo:Toper,Zelditch:Szego,Donaldson1,Donaldson2},
just to name a few.  A recent result of Donaldson
\cite{Donaldson:numeric} shows that this approximation scheme is both
mathematically elegant and relatively easy to implement
numerically.  It can also be directly generalized to find approximate
hermitian Yang-Mills connections on vector bundles \cite{en:Mike1},
which can in turn be used to compute metrics on moduli spaces, and the
kinetic terms in $N=1$ string compactifications \cite{en:Mike3}.

In this work we develop numerical methods for approximating Ricci flat
metrics on Calabi-Yau hypersurfaces based on these ideas.  This also
supplies a detailed analysis of the numerical methods used in
\cite{en:Mike1}. We study the effectiveness of our approach in the
example of a one parameter family of quintics in $\C\P^4$.  As we
review in section \ref{s:T}, we work with a space of approximating metrics
parameterized by an $N\times N$ hermitian matrix; the balanced metric
is then the fixed point of the so-called ``T map'' on this space,
defined in \eq{T}.

The main computational problem in implementing the T map numerically
is the evaluation of a large number of integrals on the manifold.
More precisely, given a Calabi-Yau $n$-fold
$X$, with its corresponding holomorphic $n$-form 
$\Omega \in
\Omega^{n,0}(X)=\Lambda^{n}(T^*X)^{1,0}$, and
volume form $d\mu_{\Omega} = \Omega\wedge
\overline{\Omega}$,
one needs to compute integrals of the type
\begin{equation}\label{e1}
\int_{X}f \; d\mu_{\Omega},
\end{equation}
where $f\colon  X\to \C$ is a smooth complex
valued (but not holomorphic) function.
Consequently, the heart of the paper (section 3) will
be devoted to developing a numerical
approximation scheme to efficiently
and accurately compute these integrals.

A second technical point, which is very valuable in simplifying these
computations, is to take advantage of the discrete symmetries of the
manifold.  We discuss this in section \ref{s:sym}.

Our explicit numerical results appear in section \ref{s:NR}, where
we also provide a general discussion of the
efficiency and accuracy of the algorithm, comparisons with alternatives,
and suggestions for future work.

Before we begin, let us briefly set out the problem.
Denote the Ricci flat metric on $X$ (which is unique 
given a complex structure
and K\"ahler class) as $g_{RF}$.  We want to propose a set of
approximating metrics $g_h$ parameterized by  parameters $h$, and
give a numerical procedure to 
find the ``best'' approximation to $g_{RF}$ within this set.

The criteria that a best approximation should satisfy include
\begin{enumerate}
\item {\bf Accuracy}: we want to minimize the error
$\epsilon=d(g_h,g_{RF})$, where $d$ is some measure of the
distance between the approximate and true metrics.
A simple and natural choice for $\epsilon$
in the present context is to consider the function
\begin{equation}\label{define-eta}
\eta_h = \frac{\det \omega_h}{\Omega\wedge\bar\Omega}
\end{equation}
on $X$, where $\omega_h$ is the K\"ahler form for $g_h$.
For a Ricci flat metric, this will be the  constant function.  We then take \footnote{
We consider only compact varieties, hence both the minimum and 
the maximum are attained.}
\begin{equation}\label{define-error}
\epsilon = 1-\frac{{\rm min}_{x\in X}~\eta_h(x)}{{\rm max}_{x\in X}~\eta_h(x)} .
\end{equation}
Of course, one could use other norms, such as 
$||\eta_h-\frac{1}{{\rm vol}~X}\int\eta_h||_p$, or curvature integrals.
\item {\bf Control}: we want an explicit bound on the error,
\begin{equation}\label{control-eta}
\epsilon(g_h,g_{RF}) < \epsilon_{max},
\end{equation}
depending on the parameters of the problem.
\item {\bf Systematic improvement}: we would like to have a control
parameter $k$, such that by increasing $k$, we can bring the error estimate
$\epsilon_{max}$ down to any desired accuracy.
\item {\bf Mathematical naturalness}.
Our experience with string theory (and more
generally in mathematics and physics) 
has been that in exploratory work such as this, 
rather than trying to incorporate all known aspects of a problem and
find a ``best'' solution, we can learn far more 
by studying a well chosen simplification in depth.  This favors a scheme
in which one makes the smallest possible number of arbitrary or
{\it ad hoc} choices not inherent in the original statement of the problem.
\end{enumerate}
Of course the approximation
should be efficiently computable as well.  We will comment
on these various aspects as they arise.

\section{T-map}    \label{s:T}

In  this section we review the construction
of balanced metrics, which in turn lead to a
convenient approximation of the Ricci flat
metric on Calabi-Yau threefolds. Our
presentation is based on \cite{en:Mike1},
which in turn was deeply inspired by
\cite{Donaldson:numeric}. We refer to these
papers, and the references therein, for more
details.

We start with a holomorphic line bundle $\cL$
on a K\"ahler manifold $X$, 
with $N$ global sections. This fact is
usually phrased as $\H^{0}(X,\,
\mathcal{L})=\C^N$. 
Let $\{s_\alpha\}_{\alpha=1}^N$ be a basis of
this vector space, and consider the map
$$
i\colon X\longrightarrow \cp{N-1}
 \qquad i(Z_0,\ldots,Z_n) =
(s_1(Z),s_2(Z),\ldots,s_N(Z)).
$$
The geometric picture is that each point in
our original manifold $X$
(with coordinates $Z_i$)
corresponds to a point in $\bbc^{N}$
parameterized by the sections
$s_\alpha$.  Since choosing a different frame
for $\cL$ would produce
an overall rescaling $s_\alpha\rightarrow
\lambda s_\alpha$, the overall
scale is undetermined. Granting that 
$s_1(Z),s_2(Z),\ldots,s_N(Z)$ 
do not vanish simultaneously, this gives us a
map into $\P^{N-1}$. 

We want this map to be an
embedding, {\it i.e.} that distinct points on $X$
map to distinct points on $\mathbb{P}^{N-1}$, and that tangent vectors
are also separated.  In general, we can
appeal to the Kodaira embedding
theorem, which asserts that for $\cL$ ample
this will be true for all powers $\cL^k$,
starting with some $k_0$.  As an example, for
non-singular quintics in $\P^4$, $\mathcal{O}_{X}(k)$
is both ample and very ample for all $k\ge
1$.  

Next we consider the $N^2$-parameter family  of  K\"ahler
potentials on $\cp{N-1}$, 
\be\label{Ks}
K_{h} = \log \left(\sum_{\alpha,\bbeta}
h^{\alpha\bbeta}
 s_\alpha \bs_\bbeta \right)\equiv \log
||s||^2_h\, ,
\ee
parametrized by the $N$ by $N$ hermitian matrix $h$.  These
give rise to a family of K\"ahler metrics on $X$
by restriction, and
we will seek a ``best'' approximation to the Ricci flat metric
within this space of approximating metrics.
Note that it is the
inverse of $h$, $h^{\alpha\bbeta}$, that
appears in \eqref{Ks}. The reason for this will become clear
shortly.

Mathematically, the simplest interpretation
of \eqref{Ks} is that it
defines a hermitian metric on the line bundle
$\cL $.  This is a sesquilinear map 
from $\bar\cL\otimes \cL$ to smooth functions
$C^\infty(X)$, here defined by
$$
(s,s') = e^{-K_{h}}\cdot\bar s\cdot s' =
 \frac{\bar s\cdot s'}{\sum_{\alpha,\bbeta}
h^{\alpha\bbeta}
 s_\alpha \bs_\bbeta} .
$$
Notice that a change of frame, which acts on
our sections
as $s_\alpha\rightarrow \lambda s_\alpha$,
cancels out in this expression.

This metric allows us to define an inner
product between the global sections:
\begin{equation}\label{T1}
 \vev{s_\beta|s_\alpha} = \int_X 
\frac{s_{\alpha}\bs_{\bbeta}}{||s||^2_h} \;
d{\vol_X}.
\end{equation} 
Note that this  inner product  depends on $h$
in a nonlinear way, since $h$ appears in the
denominator, and $||s||_h$ involves the
inverse of $h$.
Here $d\vol_X$ is a volume form on $X$, which
has to be chosen.  

One choice which makes sense for any $X$ is
$d{\vol_X}=\det\omega_h$, the standard volume form $\sqrt{\det g_h}$.
For $X$ a Calabi-Yau variety, we can instead use
$d{\vol_X}=\nu= \Omega\wedge \bar{\Omega}$. The latter is 
significantly simpler, mainly
because it is $h$-independent. Our numerical procedure will use a discrete
approximation to this volume form.

The explicit form of the expression in
\eqref{T1} suggests that one studies the map
\begin{equation}\label{T} 
T(h)_{\alpha\bbeta} = \frac{N}{\vol(X)}\int_X
\frac{s_\alpha
\bs_\bbeta}{||s||^2_h}\, {d\vol_X},
\end{equation} 
dubbed the ``T-map'', which acts on the
space of hermitian matrices  in a non-linear
way.

A fixed point of this map, $T(h) = h $,
the pair $(h,s_{\alpha})$ is called a {\em balanced} embedding, and the 
metric on $X$
associated to the corresponding K\"ahler
potential \eqref{Ks} is called the balanced
metric.\footnote{In a more precise nomenclature, which we will
not use here, the term ``balanced embedding'' is reserved for the
definition with $d{\vol_X}=\det\omega_h$, while
the alternate definition with $d{\vol_X}=\nu$
is referred to as the ``$\nu$-balanced embedding.''  
Also, there are several other equivalent ways of
defining the notion of balanced metric, which
are closely linked to the notion of stability
in Mumford's sense, but this utilitarian
definition will suffice for our purposes. For
reviews of this topic we recommend 
\cite{Thomas:GITrev,Mabuchi}.}
As it turns
out, the balanced metric is unique,
up to $U(N)$ transformations of the basis of sections and
rescaling, provided that the manifold in
question has no continuous symmetries. This
is certainly the case for a \CY\ variety, and
in particular our quintics.

It turns out that the T-map is contracting, so
the simplest way to find a fixed point of the
T-map is to iterate it. 
We have the following
\begin{theo} (see, e.g. ,\cite{Sano,Donaldson1};
and \cite{Bourguignon:Yau,Donaldson:numeric} for the $\nu$-balanced case)
Suppose that $\Aut(X,L)$ is discrete. If a
balanced embedding
exists, then, for any initial hermitian
matrix $G_0$,  as
$r\to \infty$ the sequence $T^r(G_0)$
converges to a fixed point.
\end{theo}

The importance of balanced metrics stems from
a theorem that goes back to Tian
\cite{Tian:metrics} and Zelditch
\cite{Zelditch:Szego}.
Let us consider the sequence of balanced metrics associated to
the bundles ${\mathcal L}^k$ (defined in terms 
of the Fubini-Study metric $\omega_k^{FS}$ from \eq{Ks}):
\begin{equation}
\omega_k=\frac{1}{k}\, i_k^*(\omega_k^{FS}) .
\end{equation}
The rescaling is made so that 
the cohomology class of the K\"ahler form
$[\omega_k]= c_1(\cL)\in
\H^2(X,\mathbb{Z})$ is independent of $k$.
With this definition one has 
\begin{theo}\label{theo2}
Suppose $\Aut(X,\cL)$ is discrete and
$(X,\cL^k)$ is balanced for
sufficiently large $k$. If the metrics
$\omega_k$ converge
in the $C^\infty$ norm to some limit
$\omega_\infty$ as $k\to \infty$, then
$\omega_\infty$ is a K\"ahler metric in the
class
$ c_1(\cL)$ with {\em constant} scalar curvature.
\end{theo}

The constant value of the scalar curvature is
determined by $c_1(X)$.
In particular, for $c_1(X)=0$ the scalar
curvature is zero.  Thus,
the balanced metrics $\omega_k$, in the large
$k$ limit, converge to
the Ricci flat metric. Furthermore, in this
case the $1/k$ convergence is enhanced to a 
$1/k^2$ convergence. We will see this
explicitly in Section~\ref{s:NR}.

One may ask where the complex
structure and K\"ahler moduli enter in this setup. 
The complex structure enters implicitly through the basis
of holomorphic sections
$s_\alpha$.  As for the K\"ahler
class, this is $c_1(\cL)$.  Of course, the
Ricci flatness condition is scale
invariant, so the overall scale is
irrelevant; however the point of
this is that if $h^{1,1}>1$, then by
appropriately choosing $\cL$ we
choose a particular ray in the K\"ahler cone.  

Therefore, if we can find the unique balanced metric for a given
$\cL$, we have a candidate approximation scheme.  Let us evaluate it
by our criteria.  One great advantage is that we have a control
parameter $k$, which is easy to implement and is mathematically
natural.  The balanced metric is natural in many respects,
which should make it possible to get error bounds like
\eq{control-eta}, though this has not yet been done.

On the other hand, the balanced metric is not in general the
most accurate approximation within this class of metrics
to $g_{RF}$.  As discussed in
\cite{Donaldson:numeric}, one can find other series of metrics which
converge to $g_{RF}$ faster than $1/k^2$.  

As an illustration, once we choose our measure of accuracy
$\epsilon$, say \eq{define-error}, we can propose a simple
scheme which is guaranteed to be the most accurate possible.  It is a
two-step procedure in which we take the balanced metric as the
starting point for a numerical search in the space of parameters
$h$ for the metric which minimizes
$\epsilon$.  As we discuss a bit later, 
standard numerical optimization routines will work for this purpose
if the starting point is close enough to the actual minimum.
This would not be the most efficient possible approach; one could 
improve the efficiency by using information about the linearized
problem, as discussed in \cite{Donaldson:numeric}.

For the applications we have in mind, for example being able to detect
anomalously large or small numbers in observables (perhaps having to
do with singularities), control, mathematical naturalness and ease of
programming are more important than accuracy, and thus we stick to the
balanced metric in our present work.

As another simple physical illustration, suppose that by using the
techniques of \cite{en:Mike3} we could use these results to get
canonically normalized fields and physical Yukawa couplings in
quasi-realistic compactifications.  At this point we would probably be
much happier to have results for quark and lepton masses in a variety
of models which were guaranteed accurate to within a factor of $2$,
than to have ``probable'' $10\%$ accuracy in one model.  Of course
this is a rather long term goal, but the point should be clear.

\section{Numerical integration on \CY\
varieties}

\subsection{Basic setup}\label{s:nm}

It is clear from the outset that analytic evaluation of the
integrals appearing in the T-map \eqref{T} is not possible. On the
other hand, if the integrands are smooth and relatively slowly varying
functions, it will be possible to evaluate the integrals using Monte
Carlo methods.  This is clear for the sections themselves.  Since $h$
is positive definite, the denominator in \eqref{T} is strictly
positive, mitigating (though not eliminating) the possibility of
numerical blow-ups.

Let $X$ be a compact Calabi-Yau
$n$-fold,\footnote{Although most of what we
present generalizes to varieties other than
\CY , we restrict attention to these spaces
due to their importance in string theory. } 
with its corresponding holomorphic $n$-form 
$\Omega \in \Lambda^{n,0}(X)$. The volume
form $\Omega\wedge \overline{\Omega}$
determines a natural measure $d\mu_{\Omega}$
on $X$ in the sense that
\[
 \int_{X}f \,d\mu_{\Omega}=\int_{X}f
\;\Omega\wedge \overline{\Omega}.
\]
From now on we will not distinguish between a
top form and the associated measure.

We can use $d\mu_{\Omega}$ to measure
volumes. For an open set $\mathcal{U}\subset
X$ the indicator or characteristic function
$\mathbf{1}_{\mathcal{U}}$ is defined by
\begin{displaymath}
\mathbf{1}_{\mathcal{U}}(x) = 
\left\{ \begin{array}{lll} 
1 && \text{if $x \in \mathcal{U}$}\\
0 && \text{if $x \notin \mathcal{U}$}.
\end{array}
\right.
\end{displaymath} 
The measure of $\mathcal{U}$ is its volume
\begin{equation}\nonumber
 \mu_{\Omega}(\mathcal{U})=\int_{X}
\mathbf{1}_{\mathcal{U}}\,d\mu_{\Omega}
=\vol(\mathcal{U}).
\end{equation}

To do a Monte Carlo integration,
one would ideally like to produce samples of
points on $X$ which are uniformly 
distributed according to the measure
$d\mu_{\Omega}$. This means that for every 
sample of points $\{ q_i\in X
\}_{i=1}^{N_p}$,
the expected number of points within each
open subset $\mathcal{U}\subset X$ is
\[
\sum_{i=1}^{N_{p}} \mathbf{1}_{\mathcal{U}}(q_i) =
N_p\frac{\mu_{\Omega}(\mathcal{U})}{\mu_{
\Omega}(X)}.
\]
Using this, we can estimate integrals as finite sums:
\begin{equation}\label{e3}
\int_{X}f d\mu_{\Omega}\approx
\mu_{\Omega}(X)\frac{ 1
}{N_p}\,\sum_{i=1}^{N_p} f(q_i),
\end{equation}
The statistical error of such an
approximation is of order
${1}/{\sqrt{N_p}}$ times a quantity
proportional to the mean of the $f(q_i)$'s.
\cite{NR}.

In practice, producing samples of points which are distributed
according to the measure $\mu_{\Omega}$ is not so easy.
One way to overcome this problem is by producing
points which are uniformly distributed
according to another auxiliary measure, say
$d\mu_A$. Let us assume that $d\mu_A$ is
associated to the global top form $A$.
The ratio $\Omega\wedge \overline{\Omega}/ A$ is a
global function on $X$, which we call the {\em mass function} $m_A$.
At a point $x$ it is defined to be the ratio of the two top forms
evaluated at $x:$
\begin{equation}\label{e33}
 m_A(x) = \frac{\Omega\wedge
\overline{\Omega}(x)}{A(x)}.
\end{equation}
In general this function is neither constant nor holomorphic.

While one could use this information to generate a sample distributed
according to $d\mu_\Omega$ ({\it e.g.}, by rejection sampling or MCMC),
it is simplest to explicitly put the mass function into the integrand.
Thus,
given a sample of points distributed according to $d\mu_A$,
and the mass function, we can estimate \eq{e1} as
\begin{equation}\label{e2}
\int_{X}f d\mu_{\Omega} = \int_{X}f
\,\frac{d\mu_{\Omega}}{d \mu_{A}} \,d\mu_{A}
\approx \frac{\mu_{\Omega}(X)}{\sum
m_j}\sum_{i=1}^{N_p} f(q_i) m(q_i),
\end{equation}
The presence of the mass function increases the statistical error.
On the other hand, the generic values of our mass function are
order one, and this is a very mild penalty.

Rather than regarding the Monte Carlo as a way to estimate the
original T-map, an alternate point of view is to regard the right-hand
side of \eq{e2} as defining a new measure $\nu$ and a new T-map,
leading to a new $\nu$-balanced metric which approximates the desired
$\Omega\wedge\bar\Omega$-balanced metric.  An advantage of this point
of view is that in \cite{Donaldson:numeric} it is shown that (under a
very mild hypothesis on $\nu$) the new T-map is contracting, and the
new $\nu$-balanced metric is unique.  Thus, numerical pathologies will
not enter at this stage, provided that we use the {\em same} sample of
points throughout the computation of the balanced metric.  This is also
advantageous for efficiency reasons, so we always do this.  One can
then repeat the computation with different samples to estimate the
statistical error.

\subsection{Generating the sample}

We now discuss how to efficiently generate points according to a known
simple distribution.  In this paper we restrict to the case of $X$
a degree $d$ hypersurface in $\P^{n+1}$. For definiteness let $X$
be defined as the zero locus of the degree $d$ homogenous polynomial
$f$. The case of a complete intersection is a straightforward
extension. Our main interest will be in $d=n+2$, but we can be more
general for the time being.

First, it is easy to generate random points distributed according to
the Fubini-Study measure (for any $h$) in the ambient $\P^{n+1}$.  We
simply generate uniformly distributed points on $S^{2n+3}$, a standard
numerical problem, and then mod out the overall phase.

Using this distribution, one approach to generating points on $X$
would be to keep only the points that lie sufficiently close to $X$,
in other words satisfy the defining equation of $X$ with a given
precision, and then use a root finding method (say Newton's method) to
``flow'' down to $X$. In essence this is a rejection-type algorithm. 
We implemented this strategy, but it has some
problems.  First, it is hard to control the emerging distribution on
$X$ (this depends on details of the root finding method).  Second, it
is an order of magnitude slower than the second method we are about to
describe.

The approach we use starts by taking a pair of independently
chosen random points $(X,Y)\in \P^{n+1}\!\times\! \P^{n+1}$, which define
a random line in $\P^{n+1}$.  By Bezout's theorem, a generic complex
line in $\P^{n+1}$ intersects $X$ in precisely $d$ points, and we take
these $d$ points with equal weight.  Repeating this process $M$ times
generates some random distribution of $N_p=dM$ points.

One advantage of this approach is that finding all $d$ roots of $f(z)=0$
numerically is not much harder than finding one root.  But the main 
advantage, as we  show in Section~\ref{s:ec} 
using results by Shiffman and Zelditch on
zeroes of random sections, is that that the resulting points are
distributed precisely according to the Fubini-Study measure restricted
to $X$.  The mass function \eqref{e33} is then computable quite efficiently.

A possible disadvantage for some applications is that the resulting 
sample will have correlations between the points in each $d$-fold subset.
For our purpose of Monte Carlo integration, this is not a problem, as
\eqref{e2} is the expectation value of a function of a single random
variable, and does not see these correlations.  If one were considering
functions of several independent random variables, one would probably
want to further randomize the sequence (say by permuting points
between subsets) to remove these correlations.

\subsection{Expected values of
currents}\label{s:ec}

Let us start with a smooth compact algebraic
variety $X$, and an ample line bundle $\c L$
on $X$. As reviewed in Section~\ref{s:T},
this means that $\c L^k$ defines an embedding
$i_k$ into projective space for any $k\geq
k_0$, for some positive integer $k_0:$
\begin{equation}\label{e4}
i_k \colon X \longrightarrow\P \H^{0}(X,\,
\mathcal{L}^k)^{\ast}.
\end{equation}
The idea is to consider {\em random} global
sections of  $\c L^k$, distributed uniformly
according to a natural measure, and look at
the expected value of the zero locus that
they cut out in $X$. For this it is
convenient to use the Poincare dual
formulation, where the divisor associated to
a section becomes a form, and ask what is the
expected value of the random forms. Shiffman
and Zelditch answer this question, and the
more general one when we intersect $l$ such
divisors, in full generality using the
language of currents. This section is a brief
review of some aspects of their work
\cite{Shiffman:Zelditch1,Shiffman:Zelditch2}.
For brevity we adapt their results to fit our
needs, rather than reproducing them verbatim.

The space of global sections
$\Gamma=\H^{0}(X,\, \mathcal{L}^k)$ is a
complex vector space of dimension $d_k$. If
we choose a basis for it, then it
automatically defines a hermitian inner
product, with respect to which the basis in
question is orthonormal. Conversely, given a 
hermitian inner product $\langle \cdot, \cdot
\rangle$ on $\Gamma$, there is an orthonormal
basis $\c B=\{s_1,\ldots,s_{d_k}\}$ on
$\Gamma$. Now given $s\in\Gamma$, we can
expand it in the basis $\c B$, and the inner
product induces a complex Gaussian
probability measure on $\Gamma$:
\begin{equation}
d\gamma(s)=\frac{1}{\pi^m}\,e^{-||c||^2}d^{
d_k}c\,,\qquad \txt{where
$s=\sum_{j=1}^{d_k}c_js_j$ and 
$||c||^2=\sum_{j=1}^{d_k} |c_j|^2$ }\,.
\end{equation} 
Given a metric $h$ on the line bundle $\c L$,
as explained in \eqref{T1}, $h$  defines a
hermitian metric on $\Gamma$. This is the
inner product that we are going to use on
$\Gamma=\H^{0}(X,\, \mathcal{L}^k)$
throughout this section.

Given a random variable $Y$ on the
probability space $(\Gamma, d \gamma)$,  
the expected value of $Y$ in the
probability measure $d \gamma$ is
\begin{equation}\label{e66}
 E(Y) =\int_{\Gamma}Y d \gamma.
\end{equation}

We can think of the probability space
$(\Gamma, d \gamma)$ in a slightly different
way. Consider the unit sphere 
\[
 \c S \Gamma= \c S \H^0(X, \c L^{k})=\{s\in
\H^0(X, \c L^{k})\colon\langle s, s \rangle
=1\}.
\]
The Gaussian probability measure on $\Gamma$
restricts to the {\em uniform} measure $d
\mu$ on $\c S \Gamma$. The 
expected value of $Y|_{\c S\Gamma}$ is
$E(Y) =\int_{\c S\Gamma}Y d\mu $. On the
other hand, the uniform measure on the sphere
$\c S \Gamma$ descends to the Fubini-Study
measure on the projectivization $\P \Gamma$. 
This alternative view will be very useful later on.

If we choose a section $s\in \Gamma=  \H^0(X,
\c L^{k})$, then there is a divisor $Z_s$
associated to it, which, roughly speaking, is
the zeros of $s$ minus the poles of $s$.
Since we work with $\c L^{k}$ very ample,
$Z_s$ consists of only the zero locus of $s$.
Given the probability measure $d \gamma$ on 
$\Gamma$, we can choose  $s$ randomly, and 
ask what is the expected value of the random
variable $Z$ (defined by $s\mapsto Z_s$).
This same question can be asked in an
equivalent form using Poincare duality. The
Poincare dual of $Z_s$ is a $(1,1)$ form
$T_s$, and it is more convenient to work with forms in this 
context than to work with divisors. In general $T_s$ is not a
$C^\infty$-form on $X$, but it can be given
an explicit expression using the notion of
currents, i.e., distribution valued forms.

Currents are defined as it is customary in the
theory of distributions.\footnote{For a nice
introduction to distributions and currents 
in algebraic geometry the reader can consult
\cite{GH}, Chapter 0 resp. Chapter 3.} 
Let $\Omega^{p,q}_0(X)$ be the space of
compactly supported $C^{\infty}$
$(p,q)$-forms on $X$, and for now we assume
that $\dim X=n$. The space of
$(p,q)$-\emph{currents} is the distributional
dual of $\Omega_0^{n-p,n-q}(X)$:
$\mathcal{D}^{\, p,q}(X) =
\Omega_0^{n-p,n-q}(X)^\prime$.
An element of $\mathcal{D}^{\, p,q}(X)$ is a
linear functional on $ \Omega_0^{n-p,n-q}(X)$
which continuous in the $C^{\infty}$  norm.

The usefulness of currents in our context
stems from the fact that Poincare dual
$T_{Y}$ of an algebraic subvariety $Y$,
defined by 
$$
\int_{X}T_{Y}\wedge \alpha =
\int_{Y}\iota^{\ast}(\alpha) ,
\qquad\mbox{for any $\alpha\in \Omega_0^{\dim
Y,\dim Y}(X)$},
$$
oftentimes has an explicit form in terms of
currents ($\iota\colon Y\hookrightarrow X$ is the embedding). 
Let us focus on the case when $Y$
is a hypersurface, or more generally the zero
divisor of a section $s\in \Gamma=  \H^0(X,
\c L^{k})$. In this case the current is given
by the \emph{Poincare-Lelong} formula:
\begin{equation}\nonumber
T_{s} =
\frac{i}{\pi}\,\partial\bar{\partial}\,\log
\langle s, s \rangle\,\, \in \mathcal{D}^{\,
1,1}(X).
\end{equation}
$T_s$ is also known as the zero current of
${s}$. Thus, the {Poincare-Lelong} formula
induces a map
$$
T\colon\P \H^{0}(X,\, \c L^k) \longrightarrow
\mathcal{D}^{\, 1,1}(X),\qquad s\mapsto T_s.
$$

As discussed earlier, the Fubini-Study
measure makes $\P \H^{0}(X,\, \c L^k)$ into a
probability space, and we can also view $T$
as a random variable. Since the currents form
a linear space, we can inquire about their
expected value in this probability measure
$$
E(T) = \int_{\P \H^{0}(X,\, \c L^k)}T_s\;
d\mu_{FS}(s),
$$
As it happens oftentimes in the theory of
distributions, although $T_s$ is not a
$C^{\infty}$ form, $E(T)$ is, and we have the
following proposition:

\begin{prop} (\cite[Lemma
3.1]{Shiffman:Zelditch1}) \label{p:1}
With $X$ and $\c L$ as above, for $k$
sufficiently large so that Kodaira's map
$i_k$ associated to $\H^{0}(X,\, \c L^k)$, as
defined in \eqref{e4}, is an embedding, the
expected value of the random variable $T$
representing the zero current is
$$E (T_s) = \frac{1}{k}\, i_k^{\ast} \;
\omega_k^{FS},$$
where $\omega_k^{FS}$ is  the Fubini-Study
2-form on $\P\H^{0}(X,\, \c L^k)$, and
$i_k^{\ast}$ is pullback of forms (in this
case restriction).
\end{prop}

This result generalizes to the case when we
intersect several divisors, and this will be
the case of main interest to us. Since
intersection of subvarieties is Poincare dual
to the wedge product, there is an obvious
guess how Prop.~\ref{p:1} should generalize. 
Let $s_1,\ldots,s_m$ be sections of
$\H^{0}(X,\, \c L^k)$, and consider the zero
current $T^{(m)}_{s_1,\ldots,s_m} $
associated to the set
\[
 Z_{s_1,\ldots,s_m} = \{x\in X\colon
s_1(x)=\cdots = s_m(x)=0\}.
\]
It is obvious that if we consider any $m$
linear combinations of $s_1,\ldots,s_m$
(which are themselves linearly independent),
then they determine the same zero set
$Z_{s_1,\ldots,s_m}$. Hence
$Z_{s_1,\ldots,s_m}$ is really associated to
the $m$-plane spanned by $s_1,\ldots,s_m$ in
$\H^{0}(X,\, \c L^k)$. Therefore the
probability space is the Grassmannian of
$m$-dimensional subspaces of $\H^0(X,\c
L^{k})$, with its natural Haar measure
$d\mu_{Haar}$ (a generalization of the
Fubini-Study measure). So we can ask what is
the expected value of this zero current,
computed with $d\mu_{Haar}$.

\begin{prop} (\cite[Lemma
4.3]{Shiffman:Zelditch1}) \label{p:2}
In the notation of Prop. \ref{p:1}, the
expected value of  the zero current
$T^{(m)}$, associated to the simultaneous
vanishing of $m$ random sections of
$\H^{0}(X,\, \c L^k)$, distributed according
the Haar  measure of the corresponding
Grassmannian, is
\begin{equation}\label{e7}
E(T^{(m)}) = k^{m-1}\,\big( i_k^{\ast} \;
\omega_k^{FS}\big)^m
\end{equation}
\end{prop}

Note in particular, that unlike in Prop.
\ref{p:1},  the distribution is according to
the Haar measure, but the final result still
involves the Fubini-Study form on
$\H^{0}(X,\, \c L^k)$. This is in fact
natural, given that an $m$-tuple of sections,
each distributed according to Fubini-Study on
$\H^{0}(X,\, \c L^k)$, is the same thing as
one $m$-plane, distributed according to Haar
on the Grassmannian.

As usual, besides having expected values,
random variables also have variance. The zero
current $T^{(m)}$ is no different in this
respect. Its variance has been computed
recently in \cite[Theorem
1.1]{Shiffman:Zelditch2}. In particular, it
was shown that the ratio of the variance and
the expected value goes to zero as $k$ is
increased
$$\frac{\left[ \operatorname{Var}\big(T^{(m)}
\big) \right]^{1/2}}{E\big(T^{(m)} \big)}
\sim  k^{-\frac{ m}{2} -\frac {1}{4}}.$$

\subsection{The \CY\ case}

In this section we put all the pieces
together, and explicitly show how to build
the numerical measure
$\{ q_i\in X,\, m(q_i) \}_{i=1}^{N_p}$
introduced in Section~\ref{s:nm}. We will
focus on a smooth Calabi-Yau hypersurface $X$
in  $\P^{n+1}$ of degree $n+2$. Let $X$ be
given by the zero locus of the degree $n+2$
homogeneous polynomial $f$, and let $(Z_{0},
Z_{1},\ldots Z_{n+1})$ be the homogeneous
coordinates on $\P ^{n+1}$. We denote the
embedding by
\[
 i \colon X=\mathcal{Z}(f)\hookrightarrow\P
^{n+1}.
\]

Our approach is to generate random points on
$X$ using random lines on $\P ^{n+1}$,  by
looking at the intersection of these random
lines with $X$. We can view  a random line as
the intersection of $n$ random hyperplanes.
This allows us to compute the expected value
of the corresponding zero current using the
techniques of Section~\ref{s:ec}.

For computational purposes, designing an
algorithm to generate points on $X$ in such a
fashion is straightforward. To generate a
random line on $\P ^{n+1}$ we generate two
random points, which lie on the unit sphere
$\mathrm{S}^{2n+3}\subset\C^{n+2}$, and are
distributed uniformly on this sphere.
For instance, to generate random points 
uniformly on $\mathrm{S}^{2n+3}$ we can start
with the unit cube in $\R^{2n+4}$, i.e.,
$[-1,\,1]^{2n+4}\subset \R^{2n+4}$.
Using a good quality random number
generator
we generate an uniform distribution of points
in $[-1,\,1]^{2n+4}$. Now take {\em only}
those points which 
fall within the unit disk
$\mathrm{D}^{2n+4}$, and then project them
radially 
to the boundary $\partial \mathrm{D}^{2n+4}
=\mathrm{S}^{2n+3}$. 

The intersection of the random line with $X$
can be computed by restricting the defining
polynomial $f$ to the line.
As a result, computing the common zero locus
reduces to solving for the roots of a 
polynomial of degree $n+2$ in one variable.
We  find numerically the $n+2$ roots using
the Durand-Kerner algorithm \cite{Durand,Kerner}, 
which is a refinement of the multidimensional 
Newton's method applied to a polynomial. 
This whole approach turns out to be
very efficient  in practice, in that one can
generate a million points on a quintic in a
matter of seconds.

\subsubsection{The expected zero current}

We chose to work with the hyperplane line
bundle $\c L=\O_X(1)$ on $X$. $\O_X(1)$ is
ample, and its global sections are in one to
one correspondence with the homogeneous
coordinates of the ambient $\P ^{n+1}$
(restricted to $X$). The associated Kodaira
embedding is precisely the defining one:
\[
 i_1 \colon X\hookrightarrow\P
^{n+1}=\P\H^{0}(X,\, \O_X(1))^*.
\]
If we take $n$ sections of $\c L=\O_X(1)$,
and look at their common zero locus, then by
Bezout's theorem this is generically $n+2$
points (degenerations might occur).
Therefore, considering random $n$-tuples of
sections will give random $n+2$-tuples of
points on $X$. But now we can tell how these
points are distributed, provided that the
sections were distributed according to the
Fubini-Study measure on $\P\H^{0}(X,\,
\O_X(1))=\P ^{n+1}$. Using Prop.~\ref{p:2} we
know that the expected value of the zero
current associated to the $n+2$ points of
intersection is $\big( i_1^{\ast} \;
\omega^{FS}_{\P ^{n+1}}\big)^n$. This is an
$(n,n)$ form on $X$, and plugging it into
\eqref{e33} we obtain the mass formula
\begin{equation}\label{f2}
 m(x)=\frac{\Omega\wedge
\overline{\Omega}}{\big( i_1^{\ast} \;
\omega^{FS}_{\P ^{n+1}}\big)^n }(x).
\end{equation}

\subsubsection{The numerical mass}\label{s:nM}

Let us look at the two differential forms
involved in \eq{f2}. For this we first
choose affine coordinates
$w_{a}=Z_{a}/Z_{0}$, $i=1,2,\ldots,n+1$ on 
$\P^{n+1}$. The Fubini-Study 2-form on
$\P^{n+1}$ is
\begin{equation}\label{e6}
\omega^{\P ^{n+1}}_{FS} = \Bigg[ 
\frac{\sum \mathrm{d}w_{a}\wedge
\mathrm{d}\bar{w}_{a}}{1 + \sum
w_{a}\bar{w}_{a}} -
\frac{\big(
\sum\bar{w}_{a}\mathrm{d}w_{a}\big)\wedge
\big( \sum w_{a}\mathrm{d}\bar{w}_{a}\big)}
{\big(1 + \sum w_{a}\bar{w}_{a} \big)^{2}}
\Bigg].
\end{equation}
The pullback $\big( i_1^{\ast} \;
\omega^{FS}_{\P ^{n+1}}\big)^n
\in\Omega_0^{n,n}(X,\mathbb{Z})$ is a top
form on $X$. Let $x_1,\ldots,x_n$ be local
coordinates on $X$. Then 
\begin{equation}\label{f1}
 \big( i_1^{\ast} \; \omega^{FS}_{\P
^{n+1}}\big)_{i\bar\jmath}=
\frac{\partial w_a}{\partial x_i} \; \big(
\omega^{FS}_{\P ^{n+1}}\big)_{a\bar b}  \;
\overline{\frac{\partial w_{\bar b}}{\partial
x_{\bar\jmath}}},
\end{equation}
and $\big( i_1^{\ast} \; \omega^{FS}_{\P
^{n+1}}\big)^n$ is proportional to the
determinant of this matrix. For obvious
reasons we need to evaluate this determinant.
Let us outline how this can be done, paying
attention to some of the numerical aspects.

The idea is to choose local coordinates on
$X$ that are convenient to work with. Let us
start with the point $P$ on $X$ with
homogenous coordinates $Z_i$. To minimize the
numerical error we go to the affine patch
where $|Z_i|$ is maximal.
Without loss of generality let us assume that
this happens for $i=0$. The  affine coordinates are
$w_{a}=Z_{a}/Z_{0}$. 

Let $p$ be the affine form of $f$, i.e.,
$p(w)= f(1,w_{1}, w_{2},\ldots, w_{n+1})$.
This equation determines one of the $w_{a}$-s
in term of the others, as an implicit
function. Let us assume for the sake of this
presentation that $\partial p/ \partial
w_{n+1}(P)\neq 0$. The implicit function
theorem then tell us that in an open
neighborhood of $P$ $w_{n+1}$ is a function
of the remaining variables:
$w_{n+1}=w_{n+1}(w_{1}, w_{2},\ldots,
w_{n})$. This allows us to choose the
coordinates $w_{1},\ldots, w_{n}$
to be the local coordinates $x_1,\ldots,x_n$
on $X$.

This choice of coordinates is quite
advantageous for computing \eqref{f1}. All we
need is to compute $\partial w_{n+1}/\partial
x_i$, as $\partial w_j/\partial x_i=\delta_{i
j}$. This can be done algebraically, without
explicitly solving the $p=0$ equation.
Namely, using the fact that 
\[
 p( w_{1},\ldots, w_{n},w_{n+1}(w_{1},\ldots,
w_{n}))\equiv 0
\]
is the identically zero function, its
derivative with respect to any $w_{i}$
vanishes identically, for $i=1,\ldots, n$. As
a result we have that 
\begin{equation}\label{df}
 \frac{\partial w_{n+1}}{\partial w_i}(P)=
-\frac{\partial p}{\partial
w_i}(P)/\frac{\partial p}{\partial
w_{n+1}}(P).
\end{equation}
For  numerical stability one should always
solve for the variable for which
$|{\partial p}/{\partial w_i}(P)|$ is the largest.

The second differential form entering
\eqref{f2} is $\Omega\wedge
\overline{\Omega}$. The holomorphic $n$-form
$\Omega\in \Omega^{n,0}(X,\mathbb{C})$ can be
represented using the Poincare residue map 
\cite[Section 1.1]{GH}
\begin{equation}\label{omega}
\Omega = (-1)^{i-1}
\frac{\mathrm{d}w_{1}\wedge
\mathrm{d}w_{2}\ldots \wedge
\widehat{\mathrm{d}w_{i}}\wedge\ldots \wedge
\mathrm{d}w_{n}}{\partial p(w)/\partial
w_{i}}. 
\end{equation}
where $\widehat{\mathrm{d}w_{i}}$ means the
omission of $\mathrm{d}w_{i}$ in the wedge
product.

These explicit expressions allow us to perform integrals
numerically on elliptic curves, $K3$ surfaces and 
more interestingly,  quintic 3-folds. 

\subsubsection{Symmetries}\label{s:sym}

Suppose our Calabi-Yau $X$ is preserved as a complex manifold
by the action of a discrete group $\Gamma$.  Then a Ricci flat
metric whose K\"ahler class $\omega$ is preserved by $\Gamma$ will
also be $\Gamma$-invariant, because it is unique.  As we will see in this section,
the same statement applies to the balanced metrics as well.

A general hermitian $N$ by $N$ matrix has $N^2$
independent real coefficients.  On the other hand, if the Calabi-Yau
$X$ has discrete symmetries, then we expect to find symmetry
relations between the matrix elements of $T(h)$.
Taking advantage of these relations can drastically reduce the size of
the problem.
In this section, we argue that these symmetry relations are respected
by the balanced metric and the T-map, and explain how we used them
in the  examples of Section~\ref{s:NR}.

Next, let us review the symmetries of 
$X$ defined as a
hypersurface in  $\mathbb{P}^{n+1}$ by the 
degree $n+2$ homogenous polynomial
\be\label{defX}
f=\sum_{i=0}^{n+1}Z_{i}^{{n+2}}-(n+2)\psi
\prod_{i=0}^{n+1}Z_{i}.
\ee
Here $\psi$ controls the complex structure of the hypersurface. Using the fact that $X$
is
Calabi-Yau, the symmetry group is
finite.
To find the symmetries of $X$ we  consider two natural  group actions 
on $\mathbb{P}^{n+1}$, and impose conditions such that these group actions descend to $X$.

We start with the abelian group 
$$
\bigoplus_{i=0}^{n+1}
\mathbb{Z}_{p} \subset GL(n+2)
$$
that acts by independently rescaling the $n+2$ homogenous coordinates $Z_{i} \mapsto { \alpha_{i}}
Z_{{i}}$, where $\alpha_{i}$ are  $p$th roots of unity. Since the projective coordinates are defined only up to overall rescaling, we have to mod out by the diagonal action ${\triangle\mathbb{Z}_{p}} $ and  find the group
$$
\bigoplus_{i=0}^{n+1}
\mathbb{Z}_{p}/{\triangle\mathbb{Z}_{p}} \subset \mathbb{P}GL(n+2)
$$
acting on $\mathbb{P}^{{n+1}}$. 
In order for this group to descend  to $X$,
it must leave the defining equation \eqref{defX}
invariant. In the Fermat case, that is
$\psi=0$, we set $p=n+2$ and find that the
Calabi-Yau is invariant under 
\be\label{abelian}
\bigoplus_{i=0}^{n+1}
\mathbb{Z}_{n+2}/{\triangle\mathbb{Z}_{n+2}}\cong
\left( \mathbb{Z}_{n+2}\right)^{n+1}.
\ee
For non-vanishing $\psi$, the $\alpha_{i}$
have to obey the  additional constraint 
$
\prod_{i=0}^{n+1}\alpha_{i}=1.
$
This shows that the symmetry group is a subgroup of 
$\mathbb{Z}_{n+2}/{\triangle\mathbb{Z}_{n+2}}\cong
\left( \mathbb{Z}_{n+2}\right)^{n+1}$, given by the kernel of the 
product map $(\alpha_0,\ldots,\alpha_{n+1})\mapsto \prod_{i=0}^{n+1}\alpha_{i}$.
We call this group $Ab_{n+2}$, and it is clear that there is an isomorphism 
$Ab_{n+2}\cong \left( \mathbb{Z}_{n+2}\right)^n$.
For example, in the case of the torus defined by our
cubic in $\mathbb{P}^{2}$, 
$Ab_{3}\cong\mathbb{Z}_{3}$. At the
Fermat point this group is enhanced to
$\mathbb{Z}_{3}^{2}$.

The second symmetry group we consider is the
symmetric group on $n+2 $ elements $\mathbb{S}_{{n+2}}$. 
This group acts by
permuting the coordinates of
$\mathbb{P}^{n+1}$. Since \eq{defX} is
invariant under permutations, $\mathbb{S}_{n+2}$ is a symmetry
of $X$ as well.

To see how these actions on the coordinates
of $\mathbb{P}^{n+1}$ induce an action on
$\{s_{{\alpha}}\}$, the global sections of
the line bundle ${\cal L}^{k}$ on $X$ defining the
embedding in $\mathbb{P}^{N-1}$, we can use some simple algebraic geometry. 
The fact that $X$ is given by a hyperplane in
$\mathbb{P}^{n+1}$ gives a natural way to
parameterize the global sections of ${\cal
L}^{k}=\O_X(k)$. We start with the short exact sequence (SES) defining $X$:
\[
\xymatrix@1{0 \ar[r] & \O_{\P^{n+1}}(-n-2) \ar[r]^-{\cdot f} & \O_{\P^{n+1}} \ar[r] & \O_X \ar[r] & 0}
\]
Tensoring with $\O_{\P^{n+1}}(k)$, and using the fact that $\H^1(\mathbb{P}^{n+1},\mathcal{O}_{\mathbb{P}^
{n+1}}(k-n-2))=0$, for $k>0$, we get another SES:
\be\label{Mike1}
\xymatrix@1{0 \ar[r] & 
\H^{0}(\mathbb{P}^{n+1},\mathcal{O}_{\mathbb{P}^
{n+1}}(k-n-2))  \ar[r]^-{\cdot f} & 
\H^{0}(\mathbb{P}^{n+1},\mathcal{O}_{\mathbb{P}^
{n+1}}(k))  \ar[r] &  \H^{0}(X,{\cal L}^{k}) \ar[r] & 0
}
\ee
which shows that the global sections of
$\H^{0}(X,{\cal L}^{k})$ can be parameterized
by  degree $k$ monomials in $n+2$
variables modulo the ideal generated by $f$. Therefore
the sections inherit an obvious group action.

In particular, we also find that  
\[
N=\dim \H^{0}(X,{\cal L}^{k})=
\binom{n + k+1}{k}-\binom{k-1}{k-n-2}.
\]
In addition, note that the map $i_k\colon
X\hookrightarrow{\mathbb{P}}^ {N-1}$
factorizes 
\be\label{factor}
\xymatrix{*++{X} \ar@{^{(}->}[r]^-i
\ar@/_2pc/[rr]^-{i_k} & *++{\P^{n+1}} \ar@{^{(}->}[r]^-v & *++{\P^{N-1}} }
\ee
The second embedding, $v\colon\P^{n+1}
\hookrightarrow{\mathbb{P}}^ {N-1}$, is the (Veronese) map associated to
the incomplete linear system on $ \P^{n+1}$ induced by the complete 
linear system $|{\cal L}^{k}|$ on $X$.

We will now consider the consequences of these actions on the $T$-map
and the sequence of hermitian matrices $\{T^{l}(h)\}_{l=1,2,\ldots }$.
First we consider the action of $Ab_{n+2}$. We assume that $T^{0}(h)$
is invariant under the group action. This is a choice we can always
make.  Since $Ab_{n+2}$ is an abelian group, its irreducible
representations are one dimensional, and can be labeled by the
characters.  Each section $s_{\alpha}$ transforms under a character
$\chi_{{\alpha}}$. The operator $T$ is defined in terms of the
sections, and the $Ab_{n+2}$ will force some of the
$T(h)_{\alpha\bbeta}$ matrix elements to be zero. To better understand
this we look at a toy example: the integral of an odd function $a $ on
$\mathbb{R}$. The group $G$ in question is $\mathbb{Z}_2$, and acts on
$\mathbb{R}$ by $x\mapsto -x$. $\mathbb{Z}_2$ has only one nontrivial
representation, and being odd, $a $ transform in this irrep. Now we
have that
$$
\int_{-\infty}^{+\infty}a(t)\,dt=
\int_{+\infty}^{-\infty}a(-x)\,d(-x)=-\int_{
-\infty}^{+\infty}a(x)\,dx,
$$
where we used a change of variable $t=-x$. This implies that
$\int_{-\infty}^{+\infty}a(x)dx=0$. 

More generally, in $\mathbb{R}^n$ for a function $a$, a group $G$, and an element $g \in G$, we can do the change of coordinates $t=g\cdot x$ and then
\begin{equation}\label{c1}
\int_{X} a(t)\, dV(t) = \int_{g\cdot
X} a(g\cdot x)g^{{*}}\, dV(x)=
\int_{X} a(g\cdot x) \, dV(x)\,.
\end{equation}
Here we assumed the measure to be $G$-invariant.

Applying \eq{c1} for $G= Ab_{n+2}$,
and using the fact that $s_\alpha$ transform as a character of
$G= Ab_{n+2}$, it gives that
\begin{equation}
 T(h)_{\alpha\bbeta}
=\frac{N}{\vol(X)}\int_X \frac{
\chi_{{\alpha}}(u)s_\alpha
\overline{\chi_{{{\bbeta}}}(u)} \bs_\bbeta}{||s||^2_h} \, d\mu_\Omega
=\chi_{{\alpha}}(u)\overline{\chi_{\bar{\beta}}(u)}\, T(h)_{
\alpha\bbeta}.
\end{equation}
We  used the fact that $X$ is
invariant under that action of $Ab_{n+2}$,
and so is the measure $\Omega\wedge\bar{\Omega}$, and the
denominator $||s||^2_h$. (The latter follows by induction from the
initial choice of $ T^{0}(h) $ being invariant.) In particular,  if
$\chi_{{\alpha}}(u)\overline{\chi_{\bar{\beta}}(u)}\neq 1$, for any
$u\in Ab_{n+2}$,
then the corresponding $T(h)_{\alpha\bbeta}$ has
to vanish.
In our numerical routine we impose this
vanishing condition on all the matrices 
$T^{l}(h)$.

A similar argument applies for $G= \mathbb{S}_{n+2}$. 
Since $\mathbb{S}_{n+2}$ is not
abelian, and hence its generic irreducible
representations are not one dimensional, this
constraint does not result in vanishing rules, but rather sets a priori
independent coefficients of $T(h)$ equal to each other. To see how
$\mathbb{S}_{n+2}$ acts, recall from \eq{Mike1} that
$$  \H^{0}(X,{\cal L})\cong
\H^{0}(\mathbb{P}^{n+1},\mathcal{O}_{\mathbb{P}^
{n+1}}(1))=\C^{n+2}$$
is the fundamental representation of $\mathbb{S}_{n+2}$, call it $F$. Then 
$\H^{0}(\mathbb{P}^{n+1},\mathcal{O}_{\mathbb{P}^
{n+1}}(k))$ is the $k$th symmetric tensor power of $F$, 
$Sym^k\,F$, and by \eqref{Mike1} $\H^{0}(X,{\cal L}^{k})$ 
is a quotient of this. Now we can return to \eq{c1}.
Once again,
we choose $T^{0}(h)$ to be invariant under $\mathbb{S}_{n+2}$, and then 
induction and \eq{c1} tell us which matrix elements of $T(h)$
equal each other. 

Therefore, imposing the symmetries of  both
finite groups, the number of independent components of 
$T^{l}(h)$ (for any $l$) reduces significantly. To illustrate this we consider 
 $k=12$ on the quintic in $\P^4$, i.e., $n=3$ (this was the largest $k$ we computed).
In this case $N=1490$. This means that $T(h)$ is a hermitian matrix with 
2,220,100 components. Taking into account the $Ab_5$ and $\mathbb{S}_5$
relations, one is reduced to computing 9800 components. This simplification 
speaks for itself.

\section{Numerical results}\label{s:NR}

In this section we present our explicit  numerical results. 
The main object that we compute is the balanced metric
associated to the embedding of the quintic threefold
defined in \eq{defX}. For definiteness we chose to work with $\psi=0.1$, but also tested other values of $\psi$. We also considered the case of elliptic curves ($n=1$) and K3 surfaces ($n=2$). In all these cases we obtained results similar to the ones to be presented here.

To find the balanced metric we study the associated $N\times N$ matrix
$h_k$ for several values of $k$, from $k=1$  to $k=12$.
We use $h_k$ to construct the associated
K\"ahler form $\omega_k$ on $X$, and check how well it approximates
the Ricci flat metric. We do this in several ways.

First, one can study the function defined in \eq{define-eta}
$$
\eta_k = \frac{\det\,\omega_k}{\Omega\wedge\bar{\Omega}}\colon X\longrightarrow \R.
$$
For a good approximation $g_k$ to the Ricci flat metric
$g_{RF}$ the function $\eta_k$ is almost constant.
We study the behavior of $\eta_k$ statistically, by summing  over
all the regions of $X$, and also locally paying attention to certain
special regions of the threefold.

Second, one can compute the Ricci tensor of $\omega_k$.
To check pointwise how close to zero the Ricci tensor is, we need a diffeomorphism invariant
quantity. We chose to work with the Ricci scalar. We also perform this
analysis for several values of $k$, and show how the Ricci
scalars decrease pointwise with $k$.

Before presenting the results let us comment on the errors coming from
Monte Carlo integration. We estimate them by computing the balanced
metrics associated to different samples of points, and then looking at the mean and variance
of each individual matrix element.
Ideally, one would like to produce samples of points with minimal induced error.
Constructions that reduce the standard deviation of the integrals
are refinements to the theory of numerical
integration presented here. 
Markov Chain Monte Carlo techniques, construction
of lattices on Calabi-Yau varieties, and of quasi-random points on such manifolds are
different approaches that one could consider.

\subsection{Approximating volumes v.s. \CY\ volume}

Here, we consider the way the function
$$
\vert \eta_{k}-1_X\vert\colon X\longrightarrow \R_+,\qquad  x\mapsto  \vert\eta_{k}(x)-1 \vert
$$
behaves on $X$. As argued earlier, we expect $\vert \eta_{k}-1_X\vert$ to approach the constant zero function.
One can study the deviation of
$\vert \eta_{k}-1_X\vert$ from the zero function
by computing the integral
\begin{equation}\label{Aeta}
\sigma_k = \int_{X} \vert \eta_{k}-1_X\vert\,  d\mu_{\Omega}\,.
\end{equation}
We compute this integral by our Monte Carlo method, which introduces an error, and this error can be estimated by
\begin{equation}\label{SIGeta}
\delta \sigma_k = \frac{1}{\sqrt{N_p}}\left(
\int_{X} \left(\vert \eta_{k}-1_X\vert-\sigma\right)^{2}\, d\mu_{\Omega}
\right)^{1/2},
\end{equation}
where $N_p$ is the number of points used to perform the Monte Carlo
integration in \eqref{Aeta}.

\begin{figure}[h]
\begin{center}
\begin{tabular}{cc}
   \includegraphics[angle=0,width=3in]{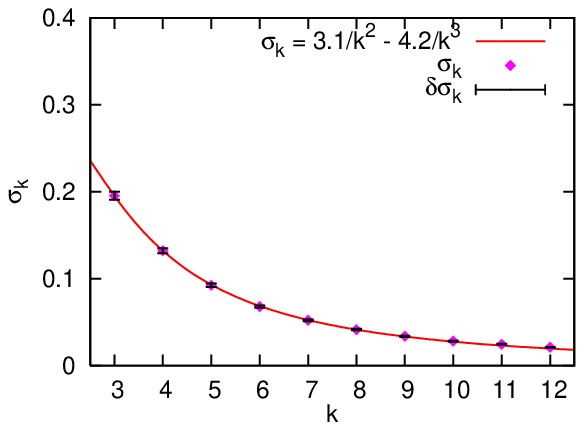}&
   \includegraphics[angle=0,width=3in]{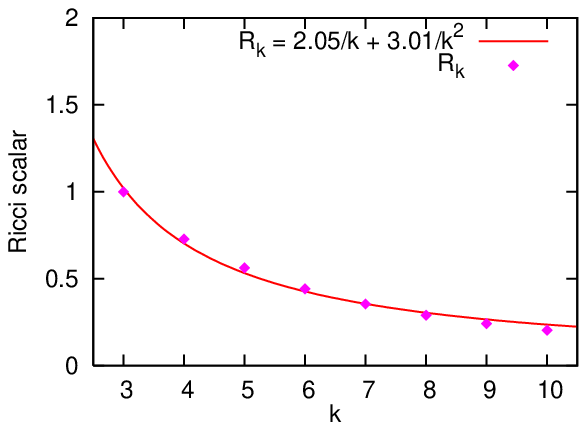}
\end{tabular}
\end{center}
    \caption{$\sigma_k$ and Ricci scalars.}	\label{f:eta}
\end{figure}

In Fig. \ref{f:eta} we plot the values  $\sigma_k$ defined in \eqref{Aeta}
for $k=3,\ldots, 12$. The error bars for each value are the
corresponding standard deviations \eqref{SIGeta}. 
We also see how the errors decrease, along with $\sigma_k$, for higher and higher $k$.
The fit in Fig. \ref{f:eta} is a curve of  type
$$
\sigma_k = \frac{\alpha}{k^2} +\frac{\beta}{k^3}  +\mathrm{O}\left(\frac{1}{k^4}\right),
$$
as we expect from the theory.

We can also study the local behavior of $\eta_k$
by restricting it to a subspace. Given our quintic 3-fold, we consider the rational curve defined by
\begin{equation}
\label{ratcuv}
( Z_0 = z_0,\, Z_1 = -z_0,\,Z_2 = z_1,\,Z_3 =
0,\,Z_4 = -z_1 ),
\end{equation}
where $Z_{i}$ are homogeneous
coordinates on $\mathbb{P}^4$, while
$(z_{0},\, z_1)$ are homogeneous coordinates
for $\mathbb{P}^1$.
This rational curve lies on every quintic
defined by \eq{defX}.

\begin{figure}[h]
\begin{center}
\begin{tabular}{cccc}
    \includegraphics[width=1.2in]{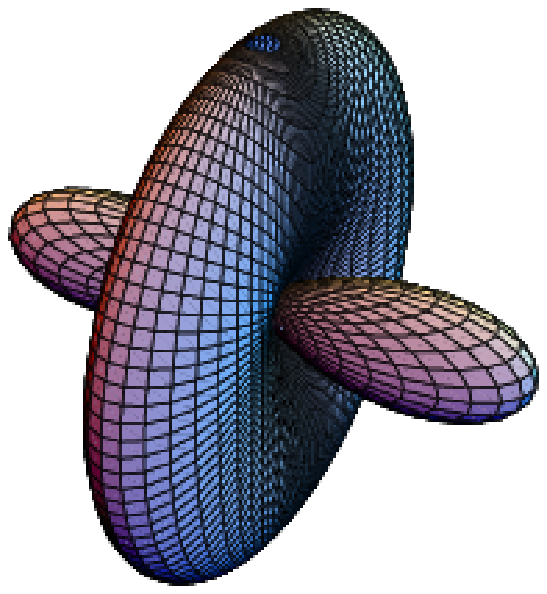}&
    \includegraphics[width=1.2in]{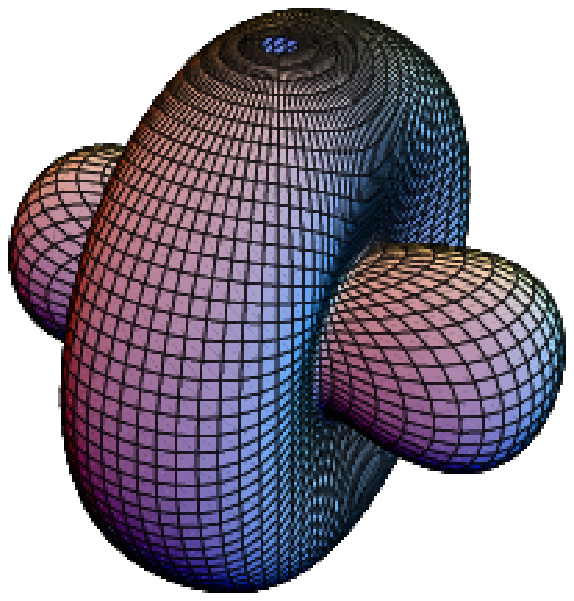}&
    \includegraphics[width=1.2in]{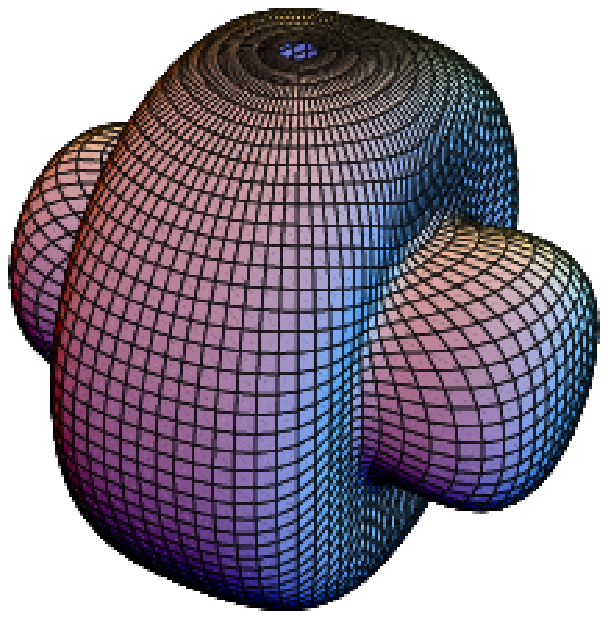}&
   \includegraphics[width=1.2in]{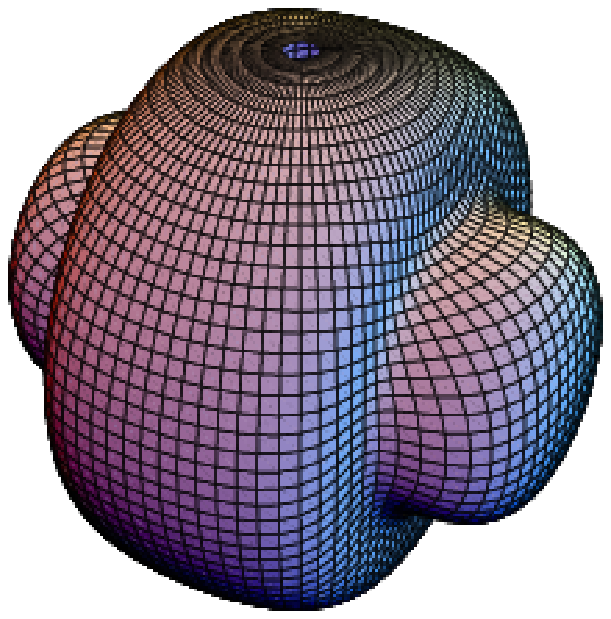}\\ \\
    \includegraphics[width=1.2in]{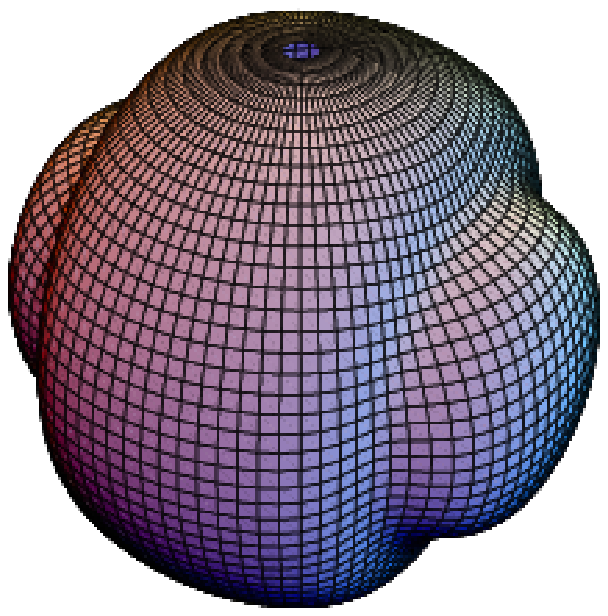}&
    \includegraphics[width=1.2in]{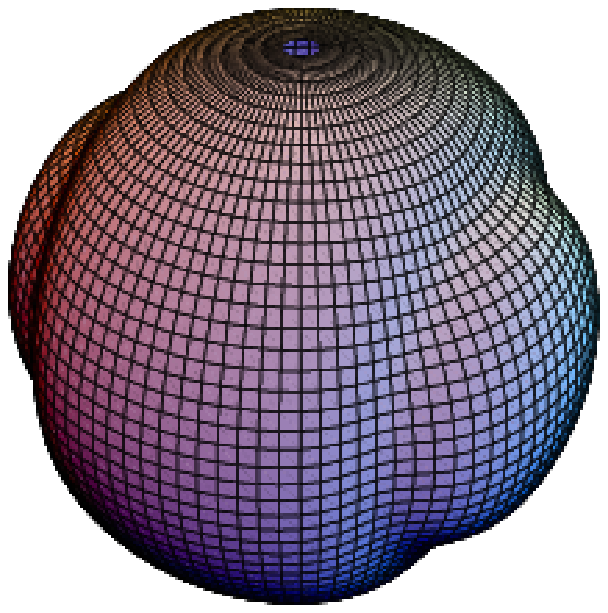}&
    \includegraphics[width=1.2in]{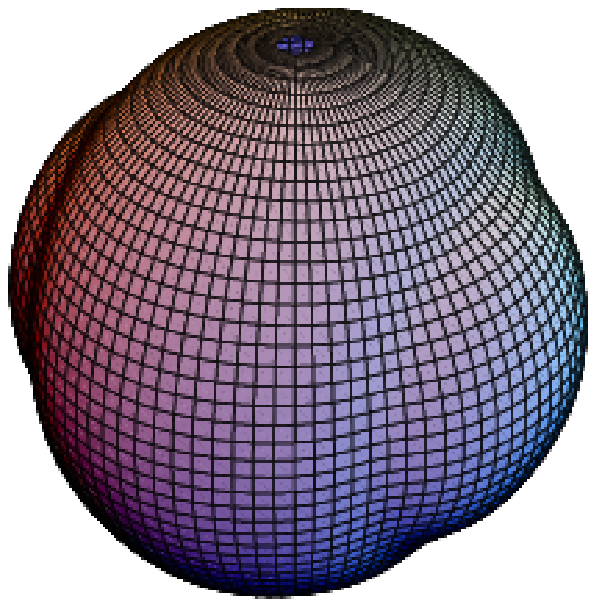}&
    \includegraphics[width=1.2in]{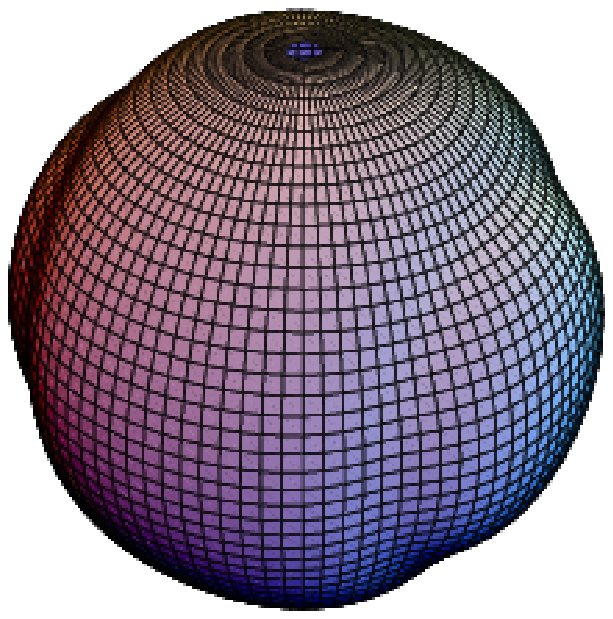}
\end{tabular}
\end{center}
    \caption{The values of $\eta$ on the rational curve, for 
$k=1$,  3, 4, 5, 7, 9, 11 and  12.}
    \label{plots}
\end{figure}

In Fig. \ref{plots} we plot the values of function $\eta_{k}$ restricted to
the rational curve defined above for 12
different values of $k$,
ranging between $1$ and $12$. 
More concretely, given the embedding \eqref{ratcuv}, we choose
the local coordinate system on $\mathbb{P}^1$
defined by
$t=z_1/z_0$, and take the stereographic
projection of
the $t$-plane. Using spherical coordinates $(\theta,\, \phi)$
on
$\P^1\equiv S^2$ we embed it into  $ \mathbb{R}^3$,
by the parameterization
$$
z_0 = \sin\theta\,\cos\phi,\qquad
z_1 = \sin\theta\,\sin\phi + i\,\cos\theta.
$$
In the radial direction of $\mathbb{R}^3$ we
plot the function $\eta_{k}$. As expected, $\eta_k$ approaches
the constant function 1 as $k$ increases.

\subsection{Ricci scalars}

Next we discuss how we compute the
Ricci scalar on $X$.
The K\"ahler potential on $X$ is given by the
restriction of the K\"ahler potential
\eqref{Ks}, which is associated to a balanced metric. The metric and Ricci curvature on
$X$ are given by
\begin{equation}\label{r4}
 g_h=\partial\bar{\partial} K_{h}
,\qquad
Ric =\partial\bar{\partial} \det{g_h}.
\end{equation}

To compute these quantities  we
need the first and second derivatives  of the sections
$s_{\alpha}$ with respect to the
coordinates of $X$. 
We can compute these derivatives algebraically,
using the ideas outlined in Section~\ref{s:nM}.
To get the actual value of
$g$ and $Ric$, at a specific point $x \in X$, we
evaluate our numerical expressions at
this point. Let us stress that this way $g$ and $Ric$
are evaluated algebraically, without numerical derivatives.

Let us now discuss the effect of the parameter $k$ on the Ricci scalar $R$. The theoretical estimate predicts a vanishing of $R$ for large $k$ as
\be\label{vanR}
||R_k|| < \frac{\gamma}{k} + \frac{\delta}{k^2}+\cdots, 
\ee
where $\gamma $ and $\delta$ are constants, in any $C^r$ norm. We can see this pointwise, after a short statistical analysis. Let us pick $100$ randomly chosen points $P_i$ on the 3-fold, and compute the associated sets of Ricci scalars $R_{k}$, for $k \in [3,\dots, 10]$. To obtain a normalized set $\tilde{R}_{k}$ of Ricci scalars we rescale all $R_{k} $ by $1/R_{3}$, and we do this for every point. This  normalization leads to a more meaningful comparison between the different points.

We are only interested in points where $\tilde{R}_{k}$ shows a generic behaviour. For this we compute the expectation value of $|\tilde{R}_{10}|$ ($k=10$ gives a good accuracy)
$$
\langle |\tilde{R}_{10}| \rangle = \frac{1}{100}\sum_{i=1}^{100} |\tilde{R}_{{10}}(P_i)|=0.48.
$$
We consider a point $P$ generic if it lies at a distance of order one from the mean $\langle |\tilde{R}_{10}| \rangle$, that is
$$
|\tilde{R}_{{10}}(P)-\langle |\tilde{R}_{10}| \rangle |<0.5.
$$
We find that $95$ points out of $100$ obey this condition. We use these $95$ points for our statistical check of \eq{vanR}.
In Fig.~\ref{f:eta} we plot $ \langle |\tilde{R}_{k}| \rangle$ as a function of $k$. We do a least square fit for both $\gamma $ and $\delta$, and obtain very good agreement.

\begin{figure}[h]
\begin{center}
\begin{tabular}{cc}
   \includegraphics[angle=0,width=3in]{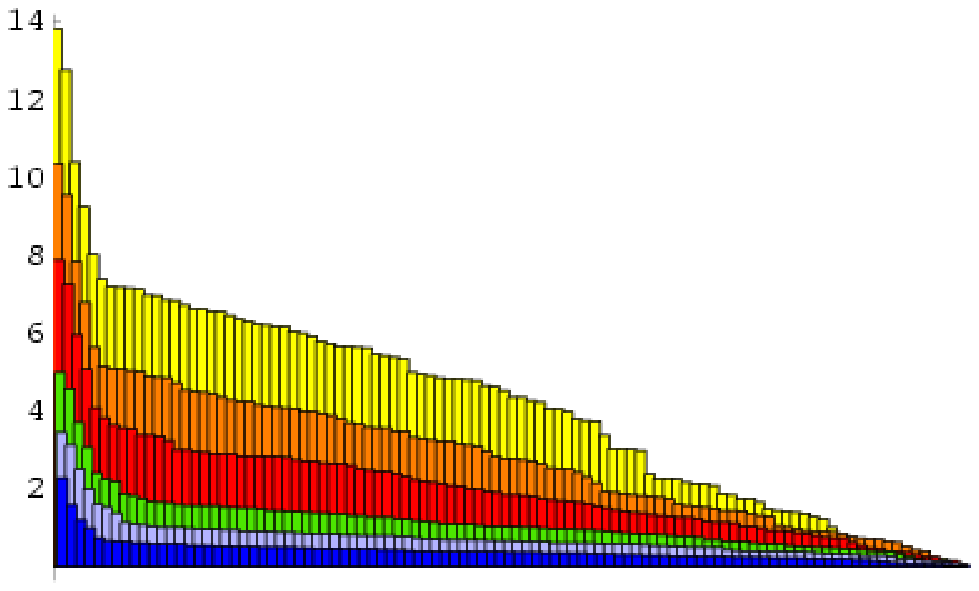}&
    \includegraphics[angle=0,width=3in]{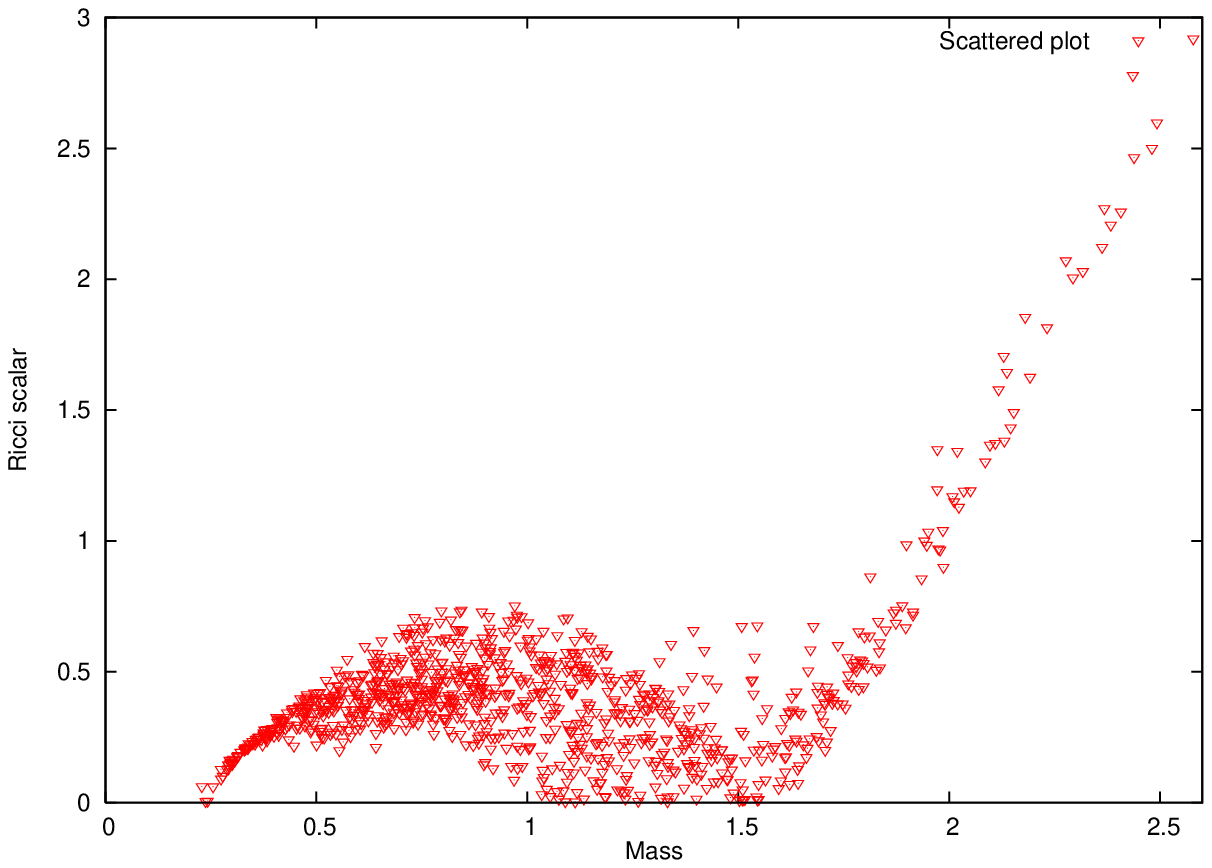}
\end{tabular}
\end{center}
    \caption{Histogram of Ricci scalars, and scattered plot of mass versus the Ricci scalars.}	
    \label{f:Histogram}
\end{figure}

We can obtain a visual picture of how the Ricci scalars decrease with $k$ by plotting them on a histogram. We use the same $100$ points from above. To keep the picture simple we plot only six $k$ values: 3, 4, 5, 7, 9 and 12. Unlike earlier, here we do not normalize the Ricci scalars pointwise, instead we reorder them decreasingly. This ordering is done for no other reason but to enhance the visual clarity. Accordingly, in the first graph of Fig.~\ref{f:Histogram} we plotted the value of the Ricci scalar for every point (the $y$ axis is the absolute value of Ricci scalar, while on the $x$ axis we have the points from 1 to 100). We did this for the $k$ values indicated above, and used different colors to distinguish them (e.g., yellow corresponds to $k=3$, while $k=12$ is blue). It is evident from this graph that by going from $k=3$ to $k=12$ the Ricci scalars decrease by an order of magnitude, in line with the theoretical expectation.

One observes that for any $k$ there are a few anomalously large Ricci scalars. To understand this we can do a scattered plot of the mass of that point versus the Ricci scalar in question (we present this for $k=12$), using $1000$ random points. This is the second graph in Fig.~\ref{f:Histogram}. The picture shows a correlation between large values of the Ricci scalar and large values of the mass (large in a logarithmic sense). In other words, in regions where our point generator needs large correction, via the mass, the balanced metric is a less accurate approximation of the Ricci flat metric, compared to points with smaller mass. This fact is then amplified by the formula for the Ricci tensor in \eq{r4}, where the logarithmic scale is also supplied.

Finally let us note that $1/\eta_1$ is 
precisely the mass function \eqref{e33}. This is because the $k=1$ balanced matrix is proportional to the identity, a consequence of the discrete symmetries present for our quintic. This leads to an alternative interpretation for the first graph ($k=1$) in Fig.~\ref{plots}, as depicting the inverse masses of the points on that rational curve.

\subsection{Discussion}

We will discuss further applications of these results elsewhere; here we
discuss the advantages and limitations of this approach compared to
others, for example position space methods \cite{Headrick:Wiseman}.

The runtime of a computation of the balanced metric can be approximated
as
$$
T = N_{it} \times N_p \times S^2 ,
$$
where $S$ is the number of independent sections 
(taking into account discrete symmetry),
$N_p$ is the number of points used in the Monte Carlo integration,
and $N_{it}$ is the number of iterations of the T-map
required for convergence.  Since convergence is exponential,
this leads to a rough scaling with the accuracy as
$$
T \sim \frac{\log \epsilon}{\epsilon^2} S^2 .
$$

The value of $S$ required for a given accuracy depends on the
symmetries and dimension.  For the balanced metrics, we expect
to need $k\sim 1/\sqrt{\epsilon}$; as discussed in section 2
this could probably be improved by choosing a different scheme
if accuracy were paramount.  For hypersurfaces in $n$ complex dimensions, 
we then have $S \sim N \sim k^{n+1}$, leading to a rough overall scaling
of $T \sim 1/\epsilon^{n+3}$.  This might be compared with a (naive)
$T \sim 1/\epsilon^{2n}$ for position space methods, so the two appear
generally competitive.  However, along with the other
advantages we mentioned, we believe the approach we are discussing is
far easier to program, and requires relatively little effort to adapt
to different manifolds, and related problems such as hermitian Yang-Mills.

Since the sections $s_\alpha$ of $\mathcal{O}_{X}(k)$ are degree $k$
polynomials, this basis is a simple type of Fourier or momentum space
basis.  Very roughly speaking, a degree $k$ basis should be able to
represent arbitrary structures on length scales down to $1/k$.  They
are particularly well suited for approximating smooth functions, as
the Fourier coefficients of such a function fall off faster than any
power of $k$ (see the appendix of \cite{Donaldson:numeric} for more
precise statements).  This is advantageous as the Ricci flat metric is
smooth, suggesting that other approximation schemes could do better
than $\epsilon\sim 1/k^2$.

On the other hand, in some limits (say a conifold limit) the metric
can develop structure on small scales, which might not be well
represented by a fixed $k$ basis.  This is also a problem for position
space methods with a fixed lattice; there one deals with it by multi-scale
methods, for example allowing the lattice spacing and structure to vary
over the manifold.  This is very powerful but also very intricate to 
program.  In the present context, rather than increase $k$, one might
look for analogous simplifications; either a multi-scale method which uses
different $k$ in different regions (or even some sort of wavelet-inspired
method).  Or, since we have many explicit expressions for Ricci flat
metrics near singularities, it might be useful to develop 
a way to patch these solutions into the global approximate solutions
we discussed.

\subsubsection*{On the computer code}

Our numerics is based on code that has been written entirely in C++. Our experience shows that these computations must be done in a compiled language, rather than an interpreted one. We have made extensive use of the following Boost libraries: uBlas, random, bind and thread. These libraries are on par with Fortran code, due to implementation techniques using expression templates and template metaprograms. The computations were done on an Athlon 64 4800+ dual core machine, with 4GB memory. The computational time ranges from minutes, for low $k$, to hours, and eventually 2 days (for $k=12$).

\subsubsection*{Acknowledgments}

This research was supported in part by the
DOE grant DE-FG02-96ER40949. 


\bibliographystyle{/home/karp/lat/utcaps}

\begin{thebibliography}{10}

\bibitem{Yau}
S.~T. Yau, ``On the {R}icci curvature of a compact {K}\"ahler manifold and the
  complex {M}onge-{A}mp\`ere equation. {I},'' {\em Comm. Pure Appl. Math.} {\bf
  31} (1978), no.~3, 339--411.

\bibitem{Tian:metrics}
G.~Tian, ``On a set of polarized {K}\"ahler metrics on algebraic manifolds,''
  {\em J. Differential Geom.} {\bf 32} (1990), no.~1, 99--130.

\bibitem{Bourguignon:Yau}
J.-P. Bourguignon, P.~Li, and S.-T. Yau, ``Upper bound for the first eigenvalue
  of algebraic submanifolds,'' {\em Comment. Math. Helv.} {\bf 69} (1994),
  no.~2, 199--207.

\bibitem{Luo:Toper}
H.~Luo, ``Geometric criterion for {G}ieseker-{M}umford stability of polarized
  manifolds,'' {\em J. Differential Geom.} {\bf 49} (1998), no.~3, 577--599.

\bibitem{Zelditch:Szego}
S.~Zelditch, ``Szego kernels and a theorem of {T}ian,'' {\em Internat. Math.
  Res. Notices} (1998), no.~6, 317--331.

\bibitem{Donaldson1}
S.~K. Donaldson, ``Scalar curvature and projective embeddings. {I},'' {\em J.
  Differential Geom.} {\bf 59} (2001), no.~3, 479--522.

\bibitem{Donaldson2}
S.~K. Donaldson, ``Scalar curvature and projective embeddings. {II},'' {\em Q.
  J. Math.} {\bf 56} (2005), no.~3, 345--356,
  \href{http://arXiv.org/abs/math.DG/0407534}{{\tt math.DG/0407534}}.

\bibitem{Donaldson:numeric}
S.~K. Donaldson, ``Some numerical results in complex differential geometry,''
  \href{http://arXiv.org/abs/math.DG/0512625}{{\tt math.DG/0512625}}.

\bibitem{en:Mike1}
M.~R. Douglas, R.~L. Karp, S.~Lukic, and R.~Reinbacher, ``Numerical solution to
  the hermitian Yang-Mills equation on the Fermat quintic,''
\href{http://arXiv.org/abs/hep-th/0606261}{{\tt hep-th/0606261}}.

\bibitem{en:Mike3}
M.~R. Douglas, R.~L. Karp, S.~Lukic, and R.~Reinbacher, ``The Kahler potential
  in heterotic string compactifications,''. To appear.

\bibitem{Thomas:GITrev}
R.~P. Thomas, ``Notes on GIT and symplectic reduction for bundles and
  varieties,'' \href{http://arXiv.org/abs/math.AG/0512411}{{\tt
  math.AG/0512411}}.

\bibitem{Mabuchi}
T.~Mabuchi, ``Extremal metrics and stabilities on polarized manifolds,''
  \href{http://arXiv.org/abs/math.DG/0603493}{{\tt math.DG/0603493}}.

\bibitem{Sano}
Y.~Sano, ``Numerical algorithm for finding balanced metrics,''. Tokyo Institute
  of Technology Preprint, 2004.

\bibitem{NR}
W.~H. Press, S.~A. Teukolsky, W.~T. Vetterling, and B.~P. Flannery, {\em
  Numerical recipes in {C}++}.
\newblock Cambridge University Press, Cambridge, 2002.
\newblock The art of scientific computing, 2nd edition.

\bibitem{Shiffman:Zelditch1}
B.~Shiffman and S.~Zelditch, ``Distribution of zeros of random and quantum
  chaotic sections of positive line bundles,'' {\em Comm. Math. Phys.} {\bf
  200} (1999), no.~3, 661--683, \href{http://arXiv.org/abs/math.CV/9803052}{{\tt math.CV/9803052}}.

\bibitem{Shiffman:Zelditch2}
B.~Shiffman and S.~Zelditch, ``Number variance of random zeros on complex
  manifolds,'' \href{http://arXiv.org/abs/math.CV/0608743}{{\tt
  math.CV/0608743}}.

\bibitem{GH}
P.~Griffiths and J.~Harris, {\em Principles of algebraic geometry}.
\newblock Wiley-Interscience [John Wiley \& Sons], New York, 1978.
\newblock Pure and Applied Mathematics.

\bibitem{Durand}
E.~Durand, {\em Solutions num\'eriques des \'equations alg\'ebriques. {T}ome
  {I}: \'{E}quations du type {$F(x)=0$}; racines d'un polyn\^ome}.
\newblock Masson et Cle, Editeurs, Paris, 1960.

\bibitem{Kerner}
I.~O. Kerner, ``Ein {G}esamtschrittverfahren zur {B}erechnung der {N}ullstellen
  von {P}olynomen,'' {\em Numer. Math.} {\bf 8} (1966) 290--294.

\bibitem{Headrick:Wiseman}
M.~Headrick and T.~Wiseman, ``Numerical Ricci-flat metrics on K3,'' {\em Class.
  Quant. Grav.} {\bf 22} (2005) 4931--4960,
\href{http://arXiv.org/abs/hep-th/0506129}{{\tt hep-th/0506129}}.

\end{thebibliography}
\providecommand{\href}[2]{#2}\begingroup\raggedright\endgroup

\end{document}